\documentclass[preprint,showpacs,preprintnumbers,amsmath,amssymb]{revtex4}

\usepackage{graphicx}
\usepackage{dcolumn}
\usepackage{amsmath}
\usepackage{amssymb}

\def\vk{{\bf k}}

\def\vS{{\bf S}}

\def\bra{\langle}
\def\ket{\rangle}

\newcommand{\eq}[1]{Eq.~(\ref{#1})}
\newcommand{\fig}[1]{Fig.~\ref{#1}}
\newcommand{\be}{\begin{equation}}
\newcommand{\ee}{\end{equation}}
\newcommand{\bea}{\begin{eqnarray}}
\newcommand{\no}{\nonumber}
\newcommand{\eea}{\end{eqnarray}}
\newcommand{\bean}{\begin{eqnarray*}}
\newcommand{\eean}{\end{eqnarray*}}
\newcommand{\bfi}{\begin{figure}}
\newcommand{\efi}{\end{figure}}
\newcommand{\bc}{\begin{center}}
\newcommand{\ec}{\end{center}}
\newcommand{\ba}{\begin{array}}
\newcommand{\ea}{\end{array}}

\begin{document}


\title{Effect of magnetic field on spontaneous 
Fermi surface symmetry breaking}

\author{Hiroyuki Yamase} 
\affiliation{Max-Planck-Institute for Solid State Research, 
Heisenbergstrasse 1, D-70569 Stuttgart, Germany}

\date{\today}

\begin{abstract} 
We study magnetic field effects on spontaneous Fermi surface 
symmetry breaking with $d$-wave symmetry, the so-called 
$d$-wave \lq\lq Pomeranchuk instability''. 
We use a mean-field model of electrons with a pure forward scattering 
interaction on a square lattice. 
When either the majority or the minority spin band is tuned close 
to the van Hove filling by a magnetic field, 
the Fermi surface symmetry breaking occurs in both bands, 
but with a different magnitude of the order parameter. 
The transition is typically of second order at high temperature and 
changes to first order at low temperature; 
the end points of the second order line are tricritical points. 
This qualitative picture does not change even in the limit of a 
large magnetic field, although 
the magnetic field substantially suppresses the transition 
temperature at the van Hove filling. 
The field produces neither 
a quantum critical point nor a quantum critical end point 
in our model.  
In the weak coupling limit, typical quantities characterizing 
the phase diagram have a field-independent single energy scale 
while its dimensionless coefficient varies with the field. 
The field-induced Fermi surface symmetry breaking is a promising scenario 
for the bilayer ruthenate  Sr$_{3}$Ru$_{2}$O$_{7}$,  
and future issues are discussed to establish such a scenario. 
\end{abstract}

\maketitle

\section{Introduction} 
Usually the  Fermi surface (FS) 
respects the point-group symmetry of the underlying lattice structure. 
However, recently a symmetry-breaking Fermi surface deformation  
with a $d$-wave order parameter, where the FS expands along the $k_{x}$ 
direction and shrinks along the $k_{y}$ direction, or vice versa,  was 
discussed in various two-dimensional interacting electron model 
on a square lattice, the 
$t$-$J$,\cite{yamase00,miyanaga06,edegger06} 
Hubbard,\cite{metzner00,wegner02,neumayr03} and 
extended Hubbard\cite{valenzuela01} model. 
This $d$-wave type Fermi surface deformation ($d$FSD) is often called 
$d$-wave Pomeranchuk instability, referring to Pomeranchuk's stability 
criterion for isotropic Fermi liquids.\cite{pomeranchuk58} 
However, the Fermi surface symmetry breaking can happen 
even without breaking such a criterion, since the instability is 
usually of first order at low temperature.\cite{khavkine04,yamase05} 
Moreover, the new concept of the Fermi surface symmetry breaking 
is applicable also to strongly correlated electron systems 
such as those described 
by the $t$-$J$ model.\cite{yamase00,miyanaga06,edegger06}  
The $d$FSD instability is driven by forward scattering processes 
of electrons close to van Hove points in the two-dimensional 
Brillouin zone. The instability is thus purely electronic and 
the lattice does not play a role. 
As a result, symmetry of an electronic state is reduced from $C_{4v}$ 
to $C_{2v}$,  while the lattice retains $C_{4v}$ symmetry as long as 
no electron-phonon coupling is considered. 

The $d$FSD competes with superconductivity. Several analyses of 
the Hubbard\cite{honerkamp02,kampf03} and 
$t$-$J$\cite{yamase00,edegger06} model showed that the $d$-wave 
superconductivity becomes a leading instability and the 
spontaneous $d$FSD does not happen. 
However, appreciable correlations of 
the $d$FSD remain.\cite{yamase04b} 
As a result, the system becomes very sensitive to a small external 
$xy$-anisotropy and shows a giant response to it, leading to 
a strongly deformed FS. 
This idea was invoked for high-temperature 
cuprate superconductors.\cite{yamase00} 
In particular, the recently observed anisotropy of magnetic excitations 
in YBa$_{2}$Cu$_{3}$O$_{6+x}$\cite{hinkov04} has been well 
understood in terms of $d$FSD correlations in the 
$t$-$J$ model.\cite{yamase06} 
Although the reduced symmetry due to the $d$FSD is the same as 
the electronic nematic phase proposed 
in the context of the so-called spin-charge stripes,\cite{kivelson98} 
the underlying physics is very different. 
The Fermi surface symmetry breaking does not require the assumption of 
charge stripes, but is driven by 
forward scattering processes of electrons.  
The $d$FSD provides an essentially different route to the nematic phase.

Although a spontaneous $d$FSD has not been proposed for cuprates 
because of the competition with the 
$d$-wave singlet pairing,\cite{yamase00,edegger06} 
the material Sr$_{3}$Ru$_{2}$O$_{7}$ turned out to have 
an interesting possibility of a spontaneous $d$FSD.\cite{grigera04} 
Sr$_{3}$Ru$_{2}$O$_{7}$ is the bilayered ruthenate with two 
metallic RuO$_{2}$ planes where Ru-ions form a square lattice. 
Unlike the corresponding single-layered material 
Sr$_{2}$RuO$_{4}$, a well-known triplet superconductor,\cite{mackenzie03} 
Sr$_{3}$Ru$_{2}$O$_{7}$ has a paramagnetic ground state.\cite{huang98} 
However, the material is close to a ferromagnetic transition, 
which was suggested by 
the strongly enhanced uniform magnetic 
susceptibility with a large Wilson ratio,\cite{ikeda00} 
uniaxial-pressure-induced ferromagnetic transition,\cite{ikeda04} 
inelastic neutron scattering,\cite{capogna03} and 
band structure calculations.\cite{hase97,singh01} 
By applying a magnetic field $h$, Sr$_{3}$Ru$_{2}$O$_{7}$ shows 
a metamagnetic transition 
at $h=h_{c}$, around which non-Fermi liquid behavior was observed in 
various quantities:  
resistivity,\cite{grigera01,perry01} 
specific heat,\cite{perry01,zxzhou04,perry05} 
thermal expansion,\cite{gegenwart06} and 
nuclear spin-lattice relaxation rate.\cite{kitagawa05} 
This non-Fermi liquid behavior was frequently 
discussed in terms of a putative metamagnetic quantum critical 
end point (QCEP), and in fact Sr$_{3}$Ru$_{2}$O$_{7}$ was 
often referred to as a 
system with a metamagnetic QCEP.\cite{grigera02} 
However, after improving sample quality, the hypothetical QCEP 
turned out to be hidden by a dome-shaped transition line of 
some ordered phase around $h_{c}$.\cite{grigera04} 
While a second order transition was speculated to occur around the 
center of the dome, a first order transition was confirmed 
at the edges of the transition 
line and was accompanied by a metamagnetic transition. 
Grigera {\it et al.}\cite{grigera04} associated this instability 
with the spontaneous $d$FSD, which turned out to be consistent 
with a large magnetoresistive anisotropy recently observed inside the 
dome-shaped transition line.\cite{borzi07} 

Any first order transition as a function of a magnetic field 
is generically accompanied by a metamagnetic transition, which follows 
from the stability of the thermodynamic potential. 
This was demonstrated in the case of the first order 
$d$FSD transition in connection with 
Sr$_{3}$Ru$_{2}$O$_{7}$.\cite{kee05}  
Quite recently we showed that the most salient features observed in 
Sr$_{3}$Ru$_{2}$O$_{7}$, not only the metamagnetic transition but also 
the phase diagram and the non-Fermi liquid like behavior 
of the uniform magnetic susceptibility and the specific heat coefficient, 
are well captured in terms of the $d$FSD instability near the van Hove 
singularity without invoking a putative QCEP.\cite{yamase07b} 
We also predicted anomalies 
associated with the $d$FSD instability 
in the temperature dependence of the magnetic susceptibility and 
the specific heat.\cite{yamase07b}

The main purpose of this paper is to expand 
our previous paper about the $d$FSD instability in the 
presence of a magnetic field.\cite{yamase07b} 
It is particularly interesting to perform a comprehensive analysis 
of magnetic field effects on the $d$FSD instability, since 
a magnetic field is often employed as a tuning parameter of a 
quantum phase transition. 
A naive question may be whether a QCEP is realized 
for the $d$FSD instability as for the 
ferromagnetic instability.\cite{binz04,belitz05} 
The phase diagram of the $d$FSD is known to be 
characterized by several universal numbers,\cite{yamase05} 
which can be compared directly  with experimental data. 
It is then interesting also how 
the universal numbers evolve in the presence of a magnetic field.

We analyze the $d$FSD instability in the charge channel 
in a one-band model on a square lattice. The model 
describes electrons interacting via 
a pure forward scattering interaction driving the $d$FSD 
in the presence of a magnetic field (Sec.~II).   
We solve this model numerically in Sec.~III and 
investigate the weak coupling limit analytically in Sec.~IV.  
In Sec.~V, we discuss the reported phase diagram for 
Sr$_{3}$Ru$_{2}$O$_{7}$\cite{grigera04} as well as 
relations to other scenarios such as 
a QCEP,\cite{millis02,grigera02,binz04} 
phase separation,\cite{honerkamp05} and 
magnetic domain formation.\cite{binz06} 
Section VI is the conclusion. 
In Appendix~A, detailed features of the $d$FSD phase diagram 
are presented. In Appendix~B, we analyze the $d$FSD 
instability in the spin channel, often called 
{\it spin-dependent Pomeranchuk instability}, which shows  
exactly the same phase diagram as in the charge channel.

\section{Model and formalism}
We investigate the $d$FSD instability in the charge channel under a 
magnetic field on a square lattice. 
The minimal model reads  
\begin{equation}
 H = \sum_{\vk,\sigma} (\epsilon_{\vk}^{0}-\mu) \, n_{\vk \sigma} + 
 \frac{1}{2N} \sum_{\vk,\sigma,\vk',\sigma'} f_{\vk\vk'} \, 
n_{\vk \sigma} n_{\vk' \sigma'} 
-h \sum_{\vk,\sigma}\sigma n_{\vk \sigma} \, 
 \label{f+h-model}
\end{equation}
where $n_{\vk \sigma} = c_{\vk \sigma}^{\dagger}c_{\vk \sigma}$ 
counts the electron number with momentum $\vk$ and spin $\sigma$; 
$c_{\vk \sigma}^{\dagger}$ ($c_{\vk \sigma}$) is an electron 
creation (annihilation) operator; $\mu$ is the chemical potential; 
$N$ is the number of lattice sites; $h$ is an effective magnetic field 
and is defined as $h=\frac{1}{2}\mathfrak{g}\mu_{B}H$ 
where $\mathfrak{g}$ is a $g$-factor, 
$\mu_{B}$ is Bohr magneton, and $H$ is a magnetic field. 
For hopping amplitudes $t$ and  $t'$ between nearest and 
next-nearest neighbors on the square lattice, respectively, 
the bare dispersion relation is given by
\begin{equation}
 \epsilon_{\vk}^{0}= -2 t (\cos k_{x}+\cos k_{y}) 
 -4 t'\cos k_{x} \cos k_{y}\, . 
\end{equation}
The forward scattering interaction driving the spontaneous $d$FSD 
has the form 
\begin{equation}
 f_{\vk\vk'} =  - g \, d_{\vk} d_{\vk'} \,,   \label{fkk}
\end{equation}
with a coupling constant $g\geq 0$ and
a $d$-wave form factor $d_{\vk} = \cos k_x - \cos k_y$.
This ansatz mimics the structure of the effective interaction in 
the forward scattering channel as obtained for the 
$t$-$J$,\cite{yamase00} Hubbard,\cite{metzner00} and extended
Hubbard\cite{valenzuela01} model. 
For $h=0$, this model and a similar model were studied 
in Refs.~\onlinecite{yamase05} and \onlinecite{khavkine04}, respectively.

We decouple the interaction by 
introducing a spin-dependent mean field  
\be
\eta_{\sigma}=-\frac{g}{N}\sum_{\vk} d_{\vk} 
\bra n_{\vk \sigma} \ket\,,
\ee
which becomes finite when the system breaks orientational symmetry 
and is thus the order parameter of the $d$FSD. 
The mean-field Hamiltonian reads 
\be
 H_{\rm MF} = \sum_{\vk,\sigma} \xi_{\vk \sigma} \, n_{\vk \sigma} 
 + \frac{N}{2g} \eta^{2}\,,
\ee
where 
\be
\xi_{\vk \sigma}=\epsilon_{\vk}^{0} + \eta d_{\vk} - \mu_{\sigma}\,.
\label{charge-dispersion}
\ee
Here the $\sigma$-summed mean filed $\eta=\sum_{\sigma} \eta_{\sigma}$  
enters $\xi_{\vk \sigma}$, and thus a finite $\eta_{\sigma}$ 
in general induces a finite $\eta_{-\sigma}$; 
the magnetic field is absorbed completely in the effective 
chemical potential $\mu_{\sigma}=\mu+\sigma h$. 
The grand canonical potential per lattice site is given by 
\be
 \omega = - \frac{T}{N} \sum_{\vk \sigma} 
\log(1+e^{-\xi_{\vk \sigma}/T}) + \frac{\eta^{2}}{2g} \; .
 \label{freeenergy} 
\ee
By minimizing \eq{freeenergy} with respect to $\eta$, 
we obtain a self-consistency equation 
\be
\eta=-\frac{g}{N}\sum_{\vk,\sigma} d_{\vk} f(\xi_{\vk \sigma})\,. 
\label{selfeq}
\ee
We consider the solution with $\eta\geq 0$, 
since the free energy \eq{freeenergy} is an even function with
respect to $\eta$. 
The self-consistency equation is written also as 
\be
\eta_{\sigma}=-\frac{g}{N}\sum_{\vk} d_{\vk} f(\xi_{\vk \sigma})\, .
\label{charge-eta}
\ee
Note that our Hamiltonian~(\ref{f+h-model}) 
does not allow momentum transfer, and thus the mean-field theory 
solves our model exactly in the thermodynamic limit.

\section{Numerical results} 
\begin{figure}
\centerline{\includegraphics[width=0.4\textwidth]{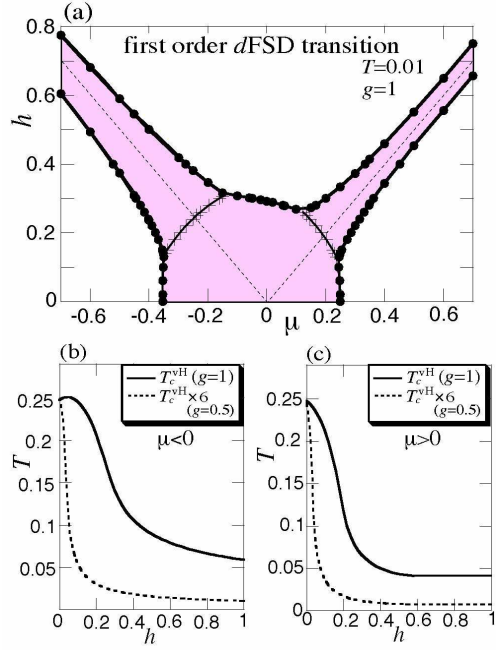}}
\caption{(Color online) 
(a) The first order $d$FSD transition line in the plane of 
$\mu$ and $h$ at $T=0.01$ for $g=1$; the symmetry-broken phase 
is stabilized in the colored (shaded) area; for notation, see the text. 
(b) and (c) $h$ dependence of $T_{c}$ along the van Hove energy, 
namely along the dotted line in (a) for $\mu<0$ and $\mu>0$, respectively; 
the solid (dotted) line is the result for $g=1$ ($0.5$); 
$T_{c}^{\rm vH}$ for $g=0.5$ is multiplied  by 6.} 
\label{phase-mh}
\end{figure}

LDA band calculations\cite{hase97,singh01} for Sr$_{3}$Ru$_{2}$O$_{7}$ 
without a magnetic field 
showed that the electronic structure is similar to that for the 
single-layered material Sr$_{2}$RuO$_{4}$ except that 
there are six FSs 
because of the bilayered structure. 
Since the $d$FSD instability is driven by electrons near the van Hove 
points on a square lattice, we focus on the FS closest to 
$\vk=(\pi,0)$ and $(0,\pi)$, and mimic such a FS by choosing $t'/t=0.35$. 
For $h=0$ the bare dispersion has the van Hove energy at $4t'=1.4t$,
from which we measure the chemical potential $\mu$.  
We take $g/t=1$ for numerical convenience, but 
the result for $g/t=0.5$ shall be 
mentioned in the context of \fig{phase-mh}. 
We set $t=1$ so that all quantities with dimension of energy are 
in units of $t$, and consider a region of $h \geq 0$ in this paper 
since the result is symmetric with respect to $h\rightarrow -h$ and 
$\sigma \rightarrow -\sigma$.

Figure~\ref{phase-m1.0}(a) shows a phase diagram for $\mu=-0.4$ 
in the plane of the magnetic field $h$ and temperature $T$; 
the dotted line at $h=h_{\rm vH}=0.4$  
represents the van Hove energy of the up-spin band (majority band).  
The $d$FSD transition is of second order for high $T$ 
and changes to first order for low $T$; 
the end points of the second order line are tricritical points. 
Hence the transition line forms  a domed shape around the van Hove 
energy. 
Figure~\ref{phase-m1.0}(b) shows the $h$ dependence of the order 
parameter $\eta$ for low $T$ together with $\eta_{\uparrow}$ 
and $\eta_{\downarrow}$. 
We see that although the phase transition happens 
around the van Hove energy of the up-spin band, 
both $\eta_{\uparrow}$ and $\eta_{\downarrow}$ show a jump 
at the first order point; 
the $\eta_{\uparrow}$ has the same sign as $\eta_{\downarrow}$, but 
with a different magnitude. 
The FSs at $T=0.01$ are shown in \fig{phase-m1.0}(c) and (d) 
for $h=0.3$ and $0.5$, respectively, together with the FSs for $g=0$;  
the outer (inner) FS corresponds to the up-spin (down-spin) 
electron band; the splitting of the FSs is due to the Zeeman energy.  
The FS instability drives a deformation of both FSs and 
typically leads to an open outer FS. 
Electron density for each spin also shows a jump at the first order 
transition point, but the size of the jump is 
different [\fig{phase-m1.0}(e)].  
This difference yields a metamagnetic transition as shown 
in \fig{phase-m1.0}(f). 
This is a generic consequence 
of the concavity of the grand canonical potential 
when a first order transition occurs as a function of $h$, 
the canonical conjugate quantity to the magnetization $m$. 
The metamagnetic transition disappears for high $T$ where the $d$FSD 
transition becomes second order.  
The $T$ dependence of the magnetization is shown in \fig{phase-m1.0}(g) 
for several choices of $h$. The magnetization shows a kink 
at the second order transition. 
Roughly speaking, compared with the non-interacting case, 
the magnetization is enhanced (suppressed) by the $d$FSD 
transition for $h \lesssim h_{\rm vH} (h \gtrsim h_{\rm vH})$. 
Since the transition is of first order for low $T$ as a function of $h$,  
there appears a phase separation as a function of the magnetization 
[\fig{phase-m1.0}(h)].  
The width of the phase separated region corresponds to a magnitude of 
a jump of $m$ at the first order transition point seen in 
\fig{phase-m1.0}(f).

The $d$FSD instability has a dome-shaped transition line around the 
van Hove energy of the up-spin band for the chemical potential ($\mu<0$)  
as shown in \fig{phase-m1.0}(a).  
When one invokes much larger $\mu(>0)$, 
the $d$FSD transition then happens around 
the van Hove energy of the down-spin band. 
When $\mu$ is around zero, 
the second order line 
extends down to $h=0$ and a first order line appears only 
on the larger $h$ side for low $T$ as shown in \fig{phase-m1.2}. 
For some special values of $\mu$, see Appendix~A.

\begin{figure}
\centerline{\includegraphics[width=0.4\textwidth]{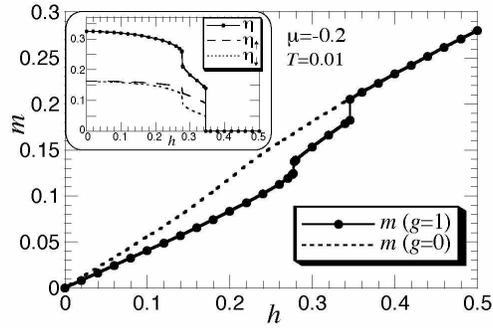}}
\caption{$h$ dependence of $m$ at $\mu=-0.2$ and $T=0.01$ for 
$g=1$ (solid line) and  $0$ (dotted line). The inset shows 
the $h$ dependence of the order parameter.}
\label{m-h-m1.2}
\end{figure}

Although a metamagnetic transition generically  
accompanies a first order $d$FSD phase transition, 
the metamagnetic transition can also occur inside the symmetry-broken phase 
because of a level crossing between two local minima of the free energy. 
This is actually the case in the present model  
as demonstrated in \fig{m-h-m1.2}, where $m$ 
is plotted as a function of $h$ for $\mu=-0.2$; 
the $h$ dependence of the order parameter is shown in the inset. 
In addition to a jump associated with the first order $d$FSD transition 
at $h\approx 0.34$ (see also \fig{phase-m1.2}), another 
metamagnetic transition appears at $h\approx 0.28$ 
in the symmetry-broken phase. 
This metamagnetic transition is a weak feature in the sense 
that it is smeared out by thermal broadening for a relatively low 
temperature while the metamagnetic transition associated with the $d$FSD 
instability from the symmetric state is robust up to $T=T_{c}^{\rm tri}$.

Figure~\ref{phase-mh}(a) summarizes numerical results  
in the plane of $(\mu,h)$ at low $T =0.01$; the solid circles denote  
a first order transition to the symmetry-broken phase, where the order 
parameter $\eta$ shows a jump from zero to a finite value, accompanied 
by a metamagnetic transition as a function of $h$; the cross symbols are 
positions, where a first order transition occurs in the symmetry-broken 
phase due to a level crossing of local minima of the free energy, 
yielding an additional metamagnetic transition; 
the dotted line represents positions of the van Hove energy for the 
up-spin band $(\mu<0)$ and the down-spin band $(\mu>0)$. 
We see that the $d$FSD phase is stabilized around the van Hove 
energy of each spin band as a function of $h$ for a given $\mu$. 
With respect to the axis $\mu=0$, the phase diagram is nearly symmetric 
and becomes fully symmetric when $t'$ is set to be zero. 
When we take a smaller $g$, the symmetry-broken phase is stabilized 
closer to the van Hove energy, 
but the qualitative features of \fig{phase-mh}(a) do not change 
except that the energy scale is substantially decreased. 

We define $T_{c}^{\rm vH}$ as $T_{c}$ at the van Hove energy, 
namely at $h=|\mu|$, 
and show its $h$ dependence in Figs.~\ref{phase-mh}(b) and (c) 
for $\mu<0$ and $\mu>0$, respectively. 
The $T_{c}^{\rm vH}$ is suppressed with increasing magnetic field. 
In particular, for a smaller $g$, 
a relatively small $h$ suppresses $T_{c}^{\rm vH}$ drastically.  
However, the suppression saturates for 
larger $h$, leading to a finite $T_{c}^{\rm vH}$, 
where we obtain a phase diagram similar to \fig{phase-m1.0}(a). 
That is, neither a quantum critical point (QCP) 
nor a QCEP of the $d$FSD is realized by the magnetic field, which 
we will further discuss in Sec.~IV.

\section{Weak coupling limit}
The $d$FSD instability occurs around the van Hove filling 
and thus the transition is dominated by states with momentum near 
the saddle points of $\epsilon_{\vk}^{0}$. 
In the weak coupling limit, 
therefore, the mean-field equations can be treated analytically 
by focusing on the state near the saddle points, 
similar to the analysis in Ref.\onlinecite{yamase05} in the 
absence of a magnetic field. 
Since the magnetic field is absorbed completely 
in the effective chemical potential $\mu_{\sigma}=\mu +\sigma h$, 
we can extend such an analysis to the 
present case by allowing the $\sigma$ dependence of the chemical potential. 
The chemical potential $\mu$ is 
measured from the van Hove energy at $h=0$ so that 
$\mu_{\sigma}=0$ indicates that the $\sigma$-spin band is 
at the van Hove filling. 
We first determine a zero temperature phase diagram 
in the plane of $(\mu,h)$, and then investigate 
$T_{c}$ suppression by a magnetic field, 
$\mu$ dependence of typical quantities characterizing 
the phase diagram, and the limit $h \rightarrow \infty$.

\subsection{Zero temperature phase diagram} 
Following the analysis in Ref.\onlinecite{yamase05}, 
the self-consistency equation \eq{selfeq} is written as 
\be
\eta= \frac{\bar g}{2}\sum_{\sigma} \big[
 (\mu_{\sigma}-\eta) \log|\mu_{\sigma}-\eta| -
             (\mu_{\sigma}+\eta) \log|\mu_{\sigma}+\eta| 
 + 2\eta\left( 1+\log\epsilon_{\Lambda} \right) \big]\,,
\label{selfeq-2}
\ee
where ${\bar g}={2m g}/{\pi^{2}}$ is the 
dimensionless coupling and 
$\epsilon_{\Lambda}=\Lambda^{2}/(2m)$ is a cutoff energy; 
$m$ is the effective mass near the van Hove energy and is 
related to the hopping integrals $t$ and $t'$.  
The grand canonical potential is then given by 
\bea
 \omega(\eta;\mu,h) &=& \frac{2m}{\pi^2} \left\{ 
 \left[ \frac{1}{2 \bar g} 
 - \frac{1}{2} - (1+\log\epsilon_{\Lambda}) \right] \eta^2 
 \right. \nonumber \\[2mm] && + \frac{1}{2}\sum_{\sigma}\left.\left[ 
 \frac{1}{2} \, (\mu_{\sigma}+\eta)^2 \log|\mu_{\sigma}+\eta| + 
 \frac{1}{2} \, (\mu_{\sigma}-\eta)^2 \log|\mu_{\sigma}-\eta| 
\right]\right\} 
 + {\rm const.} 
\label{freeenergy-2}
\eea
It is not difficult to see that 
Eqs.~(\ref{selfeq-2}) and (\ref{freeenergy-2}) 
are symmetric with respect to interchange of $h$ and $\mu$, that is,  
the magnetic field $h$ plays exactly the same role as the 
chemical potential $\mu$. We focus on the region $0 \leq \mu <h$ and 
introduce rescaled variables $\tilde{\eta}= \eta/h$ and 
$\tilde{\mu}_{\sigma} = \tilde{\mu}+\sigma$ with $\tilde{\mu}=\mu/h$. 
Equation~(\ref{selfeq-2}) then  reads 
\be
\frac{\tilde\eta}{\tilde g} = \frac{1}{2}\sum_{\sigma} \left[
 \left(\tilde{\mu}_{\sigma}-\tilde\eta\right) \log\left|\tilde{\mu}_{\sigma}-\tilde\eta\right| -
 \left(\tilde{\mu}_{\sigma}+\tilde\eta\right) \log\left|\tilde{\mu}_{\sigma}+\tilde\eta\right| \right] 
\label{selfeq-3}
\ee
with a renormalized coupling constant 
\be
 {\tilde g}^{-1} = {\bar g}^{-1} + 2\log h
 - 2(1+\log\epsilon_{\Lambda}) \; .
\label{gtilde}
\ee
Similarly \eq{freeenergy-2} is written as 
$\omega(\eta;\mu,h)=\frac{2m}{\pi^{2}}h^{2}\tilde{\omega}(\tilde{\eta};\tilde{\mu})$, where 
\bea
 \tilde\omega(\tilde\eta;\tilde{\mu})&=& 
 \left( \frac{1}{2 \tilde g} - \frac{1}{2} \right) 
 \tilde\eta^2 - \frac{1}{2}\sum_{\sigma} \tilde{\mu}_{\sigma}^{2} 
\log\left|\tilde{\mu}_{\sigma}\right| \no \\
&+& \frac{1}{2}\sum_{\sigma}\left[ 
\frac{1}{2}\left(\tilde{\mu}_{\sigma}+\tilde\eta\right)^2 
         \log\left|\tilde{\mu}_{\sigma}+\tilde\eta\right| + 
\frac{1}{2}\left(\tilde{\mu}_{\sigma}-\tilde\eta\right)^2 
           \log\left|\tilde{\mu}_{\sigma}-\tilde\eta\right|\right] 
\label{freeenergy-3}
\eea
and the energy is shifted such that 
$\tilde{\omega}(\tilde{\eta}=0;\tilde{\mu})=0$.

At zero temperature, the $d$FSD transition is usually 
of first order as we have seen in 
Figs.~\ref{phase-m1.0}(a)~and~\ref{phase-m1.2}. 
The first order transition is determined by solving 
\eq{selfeq-3} and $\tilde{\omega}(\tilde{\eta};\tilde{\mu})=0$ numerically 
for a given $\tilde{\mu}$, yielding 
a solution $\widetilde{\eta_{1}}$ and $\widetilde{g_{1}}$. 
A corresponding magnetic field $h_{1}$ is then obtained from \eq{gtilde} 
\be
h_{1}={\rm exp}\left({1+\frac{1}{2\widetilde{g_{1}}(\tilde{\mu})}}\right)
\epsilon_{\Lambda} {\rm e}^{-1/(2\bar{g})}\,,
\label{mu1}
\ee
and thus the chemical potential and the order parameter are 
$|\mu_{1}|=\tilde{\mu} h_{1}$ and 
$\eta_{1}=\widetilde{\eta_{1}}h_{1}$, respectively.   
All quantities, $h_{1}$, $\mu_{1}$, and $\eta_{1}$,  
are determined by a single energy scale 
$\epsilon_{\Lambda} {\rm e}^{-1/(2\bar{g})}$ 
and the magnetic field just changes its dimensionless coefficient. 
We plot $(\mu_{1},h_{1})$ in \fig{phase-mh-weak}; 
the phase diagram is symmetric with respect to 
the $\mu=0$ axis and the $h=|\mu|$ axis; 
such symmetry is not seen in our numerical result [\fig{phase-mh}(a)],  
since a finite $t'$ breaks the symmetry 
when $g$ is not in the weak coupling limit. 
The first order transition line has a kink at 
$|\mu| \approx 1.3 \epsilon_{\Lambda} {\rm e}^{-1/(2\bar{g})}$ and 
$h\approx 0.54 \epsilon_{\Lambda} {\rm e}^{-1/(2\bar{g})}$ 
(see the inset of \fig{phase-mh-weak}). 
At this point, the solution of 
$\tilde{\omega}(\tilde{\eta};\tilde{\mu})=0$ [\eq{freeenergy-3}] 
shows a jump, indicating 
double local minima of the free energy 
$\tilde{\omega}(\tilde{\eta})$ away from $\tilde{\eta}=0$, 
which then may yield  an additional first order transition 
in the symmetry-broken phase as seen in 
Figs.~\ref{m-h-m1.2}~and~\ref{phase-mh}(a).

\subsection{$\boldsymbol{T_{c}}$ suppression by a magnetic field} 
A magnetic field suppresses the $d$FSD transition temperature 
[\fig{phase-mh}(b)].  Here we 
clarify key factors of this suppression. 

\begin{figure}
\centerline{\includegraphics[width=0.4\textwidth]{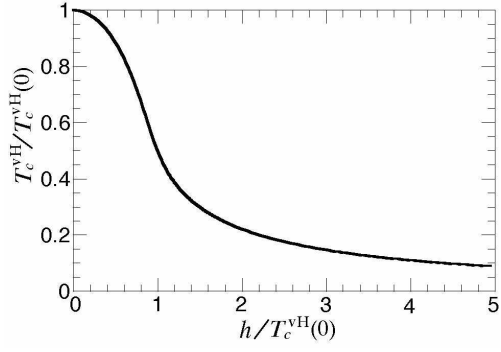}}
\caption{$T_{c}$ along the van Hove energy, namely along the 
line of $h=|\mu|$, in the weak coupling limit; $T_{c}$ and $h$ are 
scaled by $T_{c}^{\rm vH}(0)$.} 
\label{hT-weak}
\end{figure}

Since the $d$FSD transition is of second order 
as a function of $T$ at the van Hove filling, 
the $T_{c}$ is obtained by linearizing the right-hand side of \eq{selfeq} 
with respect to $\eta$, namely 
\be
1=g N_{2}(\mu,h,T_{c})\,,
\label{gapeq-2}
\ee
where we introduce 
\be
N_{p}(\mu,h,T) = - \frac{1}{N} \sum_{\vk,\sigma} d_{\vk}^{p} 
f'(\epsilon_{\vk}^{0} - \mu_{\sigma})\, ,
\label{def-Np}
\ee
a weighted density of states averaged over an energy interval of order $T$ 
around $\mu_{\sigma}$; $p$ is an even integer; $f'$ is a first 
derivative of Fermi distribution function with respect to 
$\epsilon_{\vk}^{0}$. 
In the weak coupling limit\cite{yamase05}, \eq{def-Np} reads 
\be
N_{p}(\mu,h,T) = - \frac{2m}{\pi^{2}} \sum_{\sigma} 
\int_{-\epsilon_{\Lambda}}^{\epsilon_{\Lambda}} {\rm d} \epsilon 
\log \frac{\epsilon_{\Lambda}}{|\epsilon|} f'(\epsilon- \mu_{\sigma})\,.
\label{def-Np2} 
\ee
No $p$ dependence appears on the right-hand side, since we have 
redefined $d_{\vk}=\frac{1}{2}(\cos k_{x}-\cos k_{y})$ in the present 
analysis in the weak coupling limit\cite{yamase05} 
so that $|d_{\vk}|=1$ at $\vk=(\pi,0)$ and $(0,\pi)$, 
namely at the saddle points of $\epsilon_{\vk}^{0}$. 
The van Hove filling of the $\sigma$ spin electron band is set by choosing 
\be
\mu_{\sigma}=\mu+\sigma h =0 \,, \quad 
\mu_{-\sigma}=\mu - \sigma h =-2\sigma h \,.
\label{mvH}
\ee
We consider the $\sigma = \downarrow$ case, namely for $\mu>0$ and $h>0$.  
Since 
\be
- \int_{-\epsilon_{\Lambda}}^{\epsilon_{\Lambda}} {\rm d} \epsilon 
\log \frac{\epsilon_{\Lambda}}{|\epsilon|} 
 f'(\epsilon - 2h) 
\xrightarrow{\epsilon_{\Lambda}/T \rightarrow \infty} 
\log \epsilon_{\Lambda} + a(h,T) \,,
\ee
where 
\be
a(h,T)=\int_{-\infty}^{\infty} {\rm d}x \log |Tx+2h| 
\frac{\partial}{\partial x}\frac{1}{{\rm e}^{x}+1} \,,
\ee
we obtain
\be
N_{2}(h,T) = \frac{2m}{\pi^{2}} \left[ 
2\log \frac{2\epsilon_{\Lambda}{\rm e}^{\gamma}}{\pi T} + a(h,T)-a(0,T)
\right]\,.
\label{N2-hT2}
\ee
where $a(0,T)=\log [2{\rm e}^{\gamma}/(\pi T)]$ with Euler constant 
$\gamma \approx 0.577$. 
Defining a rescaled variable $\tilde{h}_{T}=h/T$, we may write 
\bea
\zeta(\tilde{h}_{T})&=&a(0,T)-a(h,T) \no \\
&=&\log \frac{2{\rm e}^{\gamma}}{\pi}-
\int_{-\infty}^{\infty} {\rm d}x \log |x+2\tilde{h}_{T}| 
\frac{\partial}{\partial x}\frac{1}{{\rm e}^{x}+1} \,,
\label{zeta}
\eea
where $\zeta(\tilde{h}_{T})$ increases monotonically with 
$\tilde{h}_{T}$, and $\zeta(0)=0$ and 
$\zeta(\tilde{h}_{T})=\zeta(-\tilde{h}_{T})$. 
Substituting \eq{N2-hT2} into \eq{gapeq-2}, we obtain $T_{c}^{\rm vH}$ 
at the van Hove filling for a given $\tilde{h}_{T}$ 
\bea
T_{c}^{\rm vH}&=&
\frac{2 {\rm e}^{\gamma}}{\pi}\epsilon_{\Lambda}
{\rm e}^{-1/(2\bar g)}{\rm e}^{-\frac{1}{2}\zeta(\tilde{h}_{T})} 
\label{TcvH} \\ 
&=& T_{c}^{\rm vH}(0){\rm e}^{-\frac{1}{2}\zeta(\tilde{h}_{T})}
\eea
where 
\be
T_{c}^{\rm vH}(0)=\frac{2 {\rm e}^{\gamma}}{\pi}\epsilon_{\Lambda}
{\rm e}^{-1/(2\bar g)}
\label{TcvH0}
\ee
is the critical temperature for $h=0$;  
the corresponding magnetic field is obtained as 
$h=\tilde{h}_{T} T_{c}^{\rm vH}$. 
We evaluate $\zeta(\tilde{h}_{T})$ numerically and 
show in \fig{hT-weak} 
the $h$ dependence of $T_{c}^{\rm vH}$. 
The $T_{c}^{\rm vH}$ is suppressed with $h$, 
as we have seen in \fig{phase-mh}(b). 
Defining a magnetic field $h_{1/2}$, at which $T_{c}^{\rm vH}$ is 
suppressed down to a half value of $T_{c}^{\rm vH}(0)$, we 
read off $h_{1/2}=1.00 T_{c}^{\rm vH}(0)$ from \fig{hT-weak}. 
Therefore from \eq{TcvH0} we obtain 
\be
h_{1/2}\propto \epsilon_{\Lambda}{\rm e}^{-1/(2\bar g)} \,,
\label{h1/2}
\ee
that is, the suppression of $T_{c}^{\rm vH}$ is controlled by the 
dimensionless coupling constant ${\bar g}={2m g}/{\pi^{2}}$. 
When $g$ becomes smaller, $h_{1/2}$ gets 
smaller exponentially, leading to a strong suppression of $T_{c}^{\rm vH}$ 
by the magnetic field. Similarly, the smaller effective mass suppresses 
$T_{c}$ substantially with a magnetic field.

\subsection{$\boldsymbol \mu$ dependence of characteristic quantities 
of the phase diagram} 
We have derived the analytic expressions for $h_{1}$ [\eq{mu1}] and 
$T_{c}^{\rm vH}$ [\eq{TcvH}]. There is another important quantity 
characterizing the $d$FSD phase diagram, the tricritical point 
$(T_{c}^{\rm tri},h_{\rm tri})$, which 
we first calculate for a given $\mu$. 
We then summarize how $h_{1}$, $T_{c}^{\rm vH}$, $T_{c}^{\rm tri}$, 
and $h_{\rm tri}$ evolve as a function of $\mu$. 
Since all these quantities are scaled by  a single energy, 
ratios of these different quantities become universal, whose 
$\mu$ dependence is also clarified. 

\begin{figure}
\centerline{\includegraphics[width=0.4\textwidth]{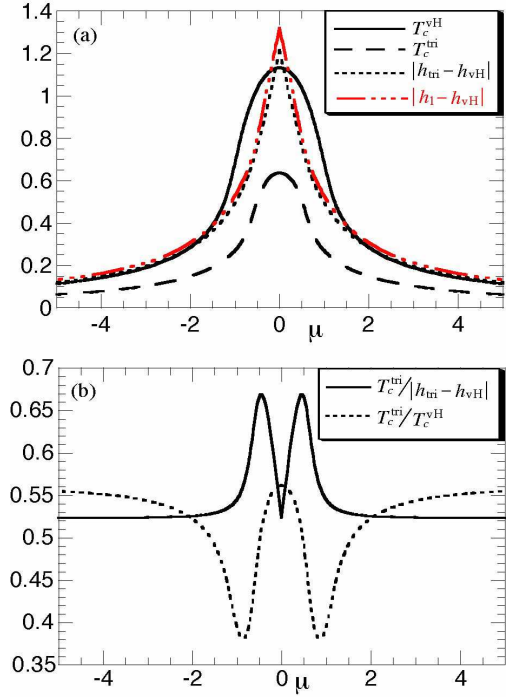}}
\caption{(Color online) 
(a) $\mu$ dependence of $T_{c}^{\rm vH}$, 
$T_{c}^{\rm tri}$, $h_{\rm tri}-h_{\rm vH}$, and $h_{1}-h_{\rm vH}$, 
where $h_{\rm vH}=|\mu|$; all quantities are scaled by the energy 
$\epsilon_{\Lambda} {\rm e}^{-1/(2\bar{g})}$. 
(b) $\mu$ dependence of the universal ratios 
$T_{c}^{\rm tri}/|h_{\rm tri}-h_{\rm vH}|$ and 
$T_{c}^{\rm tri}/T_{c}^{\rm vH}$.}
\label{Ttri-h}
\end{figure}

At the tricritical point, both the quadratic and quartic coefficient 
of the Landau free energy, $\omega(\eta)$, vanish simultaneously. 
This condition is determined by \eq{gapeq-2} and 
\be
\frac{\partial^{2}}{\partial \mu^{2}} N_{4}(\mu,h,T)=0
\label{a4zero}
\ee
where $N_{4}$ is defined in \eq{def-Np}. 
While we have analyzed \eq{gapeq-2} at the van Hove point 
[Eq.~(\ref{mvH})] in the previous section, 
we here consider \eq{gapeq-2} for general $\mu_{\sigma}$ 
and obtain 
\be
T_{c}^{\rm tri}={\rm e}^{\alpha}\epsilon_{\Lambda} 
{\rm e}^{-1/(2\bar{g})}\,,  
\label{Ttri}
\ee
where
\be
\alpha(\tilde{\mu}_{T},\tilde{h}_{T})=\frac{1}{2}\sum_{\sigma} \int_{-\infty}^{\infty}{\rm d}x \log |x+\tilde{\mu}_{T}+\sigma\tilde{h}_{T}| 
\frac{\partial}{\partial x}\frac{1}{{\rm e}^{x}+1}\,
\label{Ttri-alpha}
\ee
with $\tilde{\mu}_{T}=\mu/T$. 
Similarly \eq{a4zero} is written in the weak coupling limit as 
\be
\frac{1}{2}\sum_{\sigma} \int_{-\infty}^{\infty}{\rm d}x 
\log |x+\tilde{\mu}_{T}+\sigma\tilde{h}_{T}| 
\frac{\partial^{3}}{\partial x^{3}}\frac{1}{{\rm e}^{x}+1}=0\,.
\label{a4zero-2}
\ee
It is easy to observe that Eqs.~(\ref{Ttri-alpha})~and~(\ref{a4zero-2}) 
are symmetric with respect to 
$\tilde{\mu}_{T} \leftrightarrows \tilde{h}_{T}$ 
and thus $h$ plays exactly the same role as $\mu$. 
We solve \eq{a4zero-2} numerically for a given $\tilde{\mu}_{T}$ and  
determine $\tilde{h}_{T}$. The tricritical temperature $T_{c}^{\rm tri}$ 
is then obtained from Eqs.~(\ref{Ttri})~and~(\ref{Ttri-alpha}); 
the original $\mu$ and $h$ are determined as  
$h_{\rm tri}=\tilde{h}_{T}T_{c}^{\rm tri}$ for a given 
$\mu=\tilde{\mu}_{T} T_{c}^{\rm tri}$. 
Figure~\ref{Ttri-h}(a) summarizes $T_{c}^{\rm vH}$, $T_{c}^{\rm tri}$, 
$h_{\rm tri}$, and $h_{1}$ as a function of $\mu$. 
We see that all quantities are suppressed with increasing $|\mu|$, 
but do not reach zero; in fact they approach certain asymptotic values 
as we show in the next subsection. Since all quantities are scaled 
by the single energy, we obtain universal numbers by taking ratios of 
the different quantities. In \fig{Ttri-h}(b), we plot 
representative universal ratios 
$T_{c}^{\rm tri}/T_{c}^{\rm vH}$ and 
$T_{c}^{\rm tri}/|h_{\rm tri}-h_{\rm vH}|$, 
where $h_{\rm vH}$ is the van Hove energy and is given by 
$h_{\rm vH}=|\mu|$. 
The strong $\mu$ dependence appears for relatively small $\mu$ 
and the variation is within a factor of 1.5.

Although universal ratios are obtained in the weak coupling limit, 
these numbers also characterize well the phase diagram for 
a relatively large $g$. For example in \fig{phase-m1.0}(a), 
$T_{c}^{\rm tri}/|h_{\rm tri}-h_{\rm vH}| \approx 0.5-0.6$ and 
$T_{c}^{\rm tri}/T_{c}^{\rm vH} \approx 0.5-0.75$, 
which are comparable values to those in \fig{Ttri-h}(b).

\subsection{Large $\boldsymbol h$ limit} 
For a larger $\mu$, each quantity in \fig{Ttri-h}(a) approaches a 
certain asymptotic value. In addition, the universal ratios 
in \fig{Ttri-h}(b) converge to the same values as those at $\mu=0$. 
To understand such asymptotic behavior, we consider $T_{c}^{\rm vH}$ 
in the limit $\mu_{\uparrow}\to \infty$ while keeping $\mu_{\downarrow}=0$
[see Eq.~(\ref{mvH})], namely, the limit of $\mu \to \infty$ and 
$h \to \infty$. 
Since the up-spin electron band becomes fully occupied, we have 
$f'(\epsilon - \mu_{\uparrow})=0$ for 
$|\epsilon| <\epsilon_{\Lambda}$. 
Hence from \eq{def-Np2} we have 
\bea
N_{2}(T) &&= - \frac{2m}{\pi^{2}} 
\int_{-\epsilon_{\Lambda}}^{\epsilon_{\Lambda}} {\rm d} \epsilon 
\log \frac{\epsilon_{\Lambda}}{|\epsilon|}f'(\epsilon) \\
&&\xrightarrow{\epsilon_{\Lambda}/T\to \infty} \frac{2m}{\pi^{2}} 
\log \frac{2\epsilon_{\Lambda}{\rm e}^{\gamma}}{\pi T} \,.
\eea
The gap equation \eq{gapeq-2} then yields 
\bea
T_{c}^{\rm vH}&=&
\frac{2 {\rm e}^{\gamma}}{\pi}\epsilon_{\Lambda}
{\rm e}^{-1/\bar g}\\
&=& T_{c}^{\rm vH}(0){\rm e}^{-1/(2\bar g)}\,.
\eea
That is, the $d$FSD transition occurs even for $h\to \infty$ 
under the condition of $h=|\mu|$, although 
$T_{c}^{\rm vH}$ is suppressed a factor of ${\rm e}^{-1/(2\bar g)}$ 
compared to the case of $h=0$. 
Since the other spin band is fully occupied (empty) 
for $\mu \rightarrow + \infty$ 
($\mu \rightarrow - \infty$), 
only one spin band is subject to the $d$FSD instability. 
Therefore in the $h\to \infty$ limit, 
our model is reduced to a \lq\lq spinless'' fermion model, 
and thus the same results as those for 
$h=0$ (Sec.~V in Ref.~\onlinecite{yamase05}) 
are obtained except that the energy scale is replaced by 
$\epsilon_{\Lambda}{\rm e}^{-1/\bar g}$. 
Hence various ratios of different quantities 
characterizing the phase diagram show exactly the same 
universal numbers as those for $h=0$.  
The magnetic field just reduces the energy scale and 
cannot produce a QCP nor a QCEP of the $d$FSD instability.

\section{Relevance to $\boldsymbol {\rm Sr_{3}} \boldsymbol{\rm Ru_{2}} \boldsymbol{\rm O_{7}}$} 
We have shown that when either the up- or the down-spin electron band 
is tuned by a magnetic field close to the van Hove filling, 
the $d$FSD transition occurs in both bands but with a 
different magnitude of the order parameter. 
The field-induced $d$FSD instability is a promising scenario 
for Sr$_{3}$Ru$_{2}$O$_{7}$. 
In fact, several important features observed experimentally such as 
the phase diagram, the metamagnetic transition, and  
the non-Fermi liquid like behavior of the magnetic susceptibility 
and the specific heat coefficient are well captured in terms of 
the $d$FSD instability around 
the van Hove filling as we have shown in Ref.~\onlinecite{yamase07b}. 

On the basis of the present results, 
we discuss in more detail the reported phase diagram for 
Sr$_{3}$Ru$_{2}$O$_{7}$.\cite{grigera04} 
Since no experimental evidence of a symmetry-broken phase was 
obtained at $h=0$ and LDA band calculations\cite{hase97,singh01} 
showed that the van Hove energy is located above the Fermi energy, 
corresponding to $\mu<0$, 
we expect that the chemical potential $\mu$ is away from  
the van Hove energy, but rather close to it, 
for example $\mu\approx -0.4$ 
for the parameters shown in \fig{phase-mh}. 
Figure~\ref{phase-m1.0}(a) 
is a representative phase diagram of the magnetic 
field-induced $d$FSD instability, which is very similar to the 
reported phase diagram.\cite{grigera04} 
The maximal $T_{c}$ in the experiment is about 1 K, which is 
much smaller than the energy scale in \fig{phase-m1.0}(a). 
The coupling constant $g$ in Sr$_{3}$Ru$_{2}$O$_{7}$ is thus 
expected to be very small. 
In the weak coupling limit (Sec.~IV), 
the phase diagram is characterized by a single energy scale 
$\epsilon_{\Lambda}{\rm e}^{-1/2\bar g}$  
and thus various ratios among $T_{c}^{\rm vH}$, $T_{c}^{\rm tri}$, 
$h_{\rm tri}$, and $h_{1}$ become universal numbers [\fig{Ttri-h}(b)]. 
The universal ratios are compared directly with experimental data.
The data by Grigera {\it et al.}\cite{grigera04} 
provide $T_{c}^{\rm vH} \approx 1~{\rm K}$, 
$T_{c}^{\rm tri} \sim  0.6~{\rm K}$, 
$h_{\rm tri} \approx h_{1}=\frac{1}{2}\mathfrak{g} \mu_{B} H$ with 
$H \approx 7.8\ {\rm and}\ 8.1\ {\rm Tesla}$, and 
$h_{\rm vH}=\frac{1}{2}\mathfrak{g} \mu_{B} H_{\rm vH}$ with 
$H_{\rm vH} \approx 7.95\ {\rm Tesla}$. 
We thus obtain $h_{\rm tri}/h_{1} \approx 1$, 
$T_{c}^{\rm tri}/T_{c}^{\rm vH} \sim 0.6$, and  
$T_{c}^{\rm tri}/|h_{\rm tri}- h_{\rm vH}| \sim  (k_{B}\cdot 0.6)/
(\frac{1}{2}\mathfrak{g}\mu_{B} \cdot 0.15) \approx 12 \mathfrak{g}^{-1}$ 
with Boltzmann constant $k_{B}$.  
The first one is consistent with our result \fig{Ttri-h}(a);  
the value of the second one is comparable with \fig{Ttri-h}(b); 
as for the last one, however, the discrepancy is by a factor of 10 
if we assume $\mathfrak{g} =2$. 

While the field-induced $d$FSD instability is 
well analyzed in the present model, 
further studies are necessary 
to make more quantitative comparison with experiments. 
(i) The interaction in our model retains SU(2) symmetry 
and thus the present theory cannot address the issue why 
the possibility of the $d$FSD instability was clearly observed 
only over a narrow region of applied field angle  
to the RuO$_{2}$ plane.\cite{green05}  
One possible origin of such a field angle dependence lies in 
a magnetic anisotropy, which we expect to originate mainly from  
a relatively strong spin-orbit coupling connected with the heavy Ru-ion. 
Our model should be extended by taking the magnetic anisotropy 
into account for a more quantitative study.  
(ii) 
Inclusion of a magnetic interaction in our model may also be necessary, 
since Sr$_{3}$Ru$_{2}$O$_{7}$ is expected to be 
close to a ferromagnetic 
transition.\cite{ikeda00,ikeda04,capogna03,hase97,singh01}  
This is suggested also by comparing \fig{phase-m1.0}(g) 
with the experimental data.\cite{grigera04} 
In \fig{phase-m1.0}(g) 
the $d$FSD instability produces a kink in the $T$ dependence of $m$ 
at the transition temperature $T_{c}$. 
This kink is actually observed in 
the experiment\cite{grigera04} and 
our result for $h\gtrsim 0.35$ is similar to the experiment. 
But the present theory cannot reproduce 
the observed upward curvature of the $T$ dependence of $m$ 
for $T>T_c$,\cite{grigera04}  
which may come from a magnetic interaction.  
(iii) Sr$_{3}$Ru$_{2}$O$_{7}$ is a bilayered material. Since the $d$FSD 
instability is driven by forward scattering processes of 
quasi-particles near 
$\vk=(\pi,0)$ and $(0,\pi)$, a form of bilayer coupling 
can be important if the bilayer coupling gives rise to $k_{z}$ 
dispersion around $\vk=(\pi,0)$ and $(0,\pi)$. 
Insights into the $k_{z}$ dispersion will be obtained from further detailed 
LDA calculations.\cite{hase97,singh01}

The $d$FSD instability was discussed in basic lattice models  
such as two-dimensional 
$t$-$J$,\cite{yamase00,miyanaga06,edegger06} 
Hubbard,\cite{metzner00,wegner02,neumayr03} and 
extended Hubbard\cite{valenzuela01} model, and 
can be a generic tendency near van Hove filling in correlated electron 
systems.\cite{misctJ} 
Hence the $d$FSD is an interesting possibility when the Fermi energy 
is tuned close to the van Hove energy in other 
materials also such as Sr$_{2-y}$La$_{y}$RuO$_{4}$\cite{kikugawa04} 
where La-substitution introduces electron carriers and makes 
the FS closer to the van Hove point. 
However, La introduces some disorder in the RuO$_{2}$ plane; 
its effects should be considered carefully, since the physics near 
the van Hove singularity may in general depend strongly on sample purity. 
In fact, the specific heat coefficient for low $T$ in 
Sr$_{2-y}$La$_{y}$RuO$_{4}$\cite{kikugawa04} 
shows different behavior from that 
in Sr$_{3}$Ru$_{2}$O$_{7}$ although both systems are expected to be 
nicely tuned close to the van Hove filling. 
It is interesting to investigate impurity effects on the 
$d$FSD instability by chemical (electron) 
doping to Sr$_{3}$Ru$_{2}$O$_{7}$. 
According to our phase diagram 
[Figs.~\ref{phase-mh}(a)~and~\ref{phase-mh-weak}, see also Appendix~A], 
the $d$FSD instability can occur for a smaller magnetic field and 
even without the field if the impurity effect 
by the electron doping is not serious.

Around the van Hove filling, various ordering tendencies 
such as antiferromagnetism, ferromagnetism, superconductivity, and 
$d$-density wave develop,\cite{honerkamp02,kampf03} and 
compete with the $d$FSD instability. 
Since the $d$FSD instability is suppressed by a magnetic field 
and its suppression 
is controlled by $\overline{g} \propto mg$ [see \eq{h1/2}], 
the absolute values of the effective mass and the coupling constant are 
crucial to the possible $d$FSD instability over other instabilities. 
In this sense, microscopic derivation of $m$ and $g$ as well as 
magnetic field dependences of other instabilities are 
important future issues. 

Some order competing with the $d$FSD instability is in fact expected 
in Sr$_{3}$Ru$_{2}$O$_{7}$. 
While the experimental phase diagram\cite{grigera04} is very 
similar to our result \fig{phase-m1.0}(a), the closer comparison 
between them reveals a difference of slope of 
the first-order-transition line. In the experiment, 
the edges of the first order line are shifted to 
the center of the phase diagram 
so that the $d$FSD state is stabilized in a narrower region 
for lower $T$. 
This can be interpreted as development of some ordering tendency 
for lower $T$ in Sr$_{3}$Ru$_{2}$O$_{7}$, which then 
suppresses the $d$FSD instability.  
Although a different interpretation was given 
in Ref.~\onlinecite{grigera04}, 
a theoretical result consistent with this interpretation was indeed 
obtained in the case of 
competition of the $d$FSD and superconductivity.\cite{yamase07a}

It should be kept in mind that even if the $d$FSD instability 
does not become a leading instability, the system can still keep 
appreciable correlations of the $d$FSD.\cite{yamase04b}  
As a result, small external perturbations such as an anisotropic strain  
and a lattice anisotropy can drive sizable $d$-wave type FS deformations. 
This idea was proposed for high-temperature 
cuprate superconductors.\cite{yamase00}

Sr$_{3}$Ru$_{2}$O$_{7}$ is frequently refereed to as a material with a 
metamagnetic QCEP.\cite{grigera02} 
The compelling evidence for this was 
the systematic decrease of a metamagnetic critical end point 
by rotating the magnetic field out of the plane.\cite{grigera03misc} 
In fact, several theoretical scenarios based on a metamagnetic QCEP were 
proposed.\cite{millis02,binz04} 
However, recent data for ultrapure crystals showed 
that the critical end point does not reach zero.\cite{green05} 
Quite recently we have found that several important features are already 
well captured even without the putative QCEP.\cite{yamase07b}  
The non-Fermi liquid like behavior of 
the magnetic susceptibility\cite{ikeda00} 
and specific heat\cite{perry01,zxzhou04,perry05} 
observed in Sr$_{3}$Ru$_{2}$O$_{7}$ 
can originate from the van Hove singularity of the density of states. 
Moreover, while the first $d$FSD transition as a function of 
a magnetic field is generically accompanied by a metamagnetic transition 
[\fig{phase-m1.0}(a) and Ref.~\onlinecite{kee05}], 
the $d$FSD transition does not lead to a metamagnetic QCEP 
in the present model. 
It remains to be studied 
whether a concept of a metamagnetic QCEP can be a good basis to 
discuss electronic properties in Sr$_{3}$Ru$_{2}$O$_{7}$ and 
how the putative QCEP can be  related to the 
$d$FSD instability and the van Hove singularity. 
In this sense,
it is important to clarify whether the anomalous $T$ dependence of 
the resistivity observed around the metamagnetic transition\cite{perry01}
can be explained in terms of $d$FSD fluctuations and
the van Hove singularity
or whether 
we have to invoke quantum fluctuations originating from some QCEP.

Different scenarios from the $d$FSD and the QCEP were proposed for 
Sr$_{3}$Ru$_{2}$O$_{7}$, 
microscopic phase separation due to Coulomb energy\cite{honerkamp05} 
and magnetic domain formation due to long-range dipolar 
interactions.\cite{binz06} 
In our analysis, we employ the ground canonical ensemble 
and a first order phase transition occurs as a function of a   
magnetic field or the chemical potential for low $T$. 
Such a transition in turn appears as a magnetic phase separation 
or an electronic  phase separation as a function of its canonical 
conjugate variable, namely the magnetization [see \fig{phase-m1.0}(h)] 
or the electron 
density [see \fig{phase-m1.0}(g) in Ref.~\onlinecite{yamase05}].  
Therefore invoking the dipolar interaction or Coulomb force, we expect 
some inhomogeneous states in line with 
Refs.~\onlinecite{honerkamp05}~and~\onlinecite{binz06}. 
However our possible inhomogeneous state may replace only 
phase separated regions in \fig{phase-m1.0}(h) and thus is realized 
only near a metamagnetic transition, 
in contrast to Refs.~\onlinecite{honerkamp05}~and~\onlinecite{binz06},  
where the inhomogeneous state is stabilized in the entire region of 
the phase diagram.

\section{Conclusion}
We have performed a comprehensive analysis of 
magnetic field effects on the $d$FSD instability in a one-band 
mean-field model with a pure forward scattering interaction. 
In the plane of $\mu$ and $h(>0)$, the $d$FSD instability 
occurs around the axis of $\mu=\pm h$, namely 
around the van Hove filling of the majority band $(\mu<0)$ and 
the minority band $(\mu>0)$, even in the limit of $h \rightarrow \infty$.  
The magnetic field, however, suppresses substantially 
the energy scale of the $d$FSD instability.  
The $d$FSD instability occurs through a second order transition at  
high $T$ and typically changes to a first order transition at low $T$; 
the end points of the second order line are tricritical points. 
Neither a QCP nor a QCEP of the $d$FSD 
is realized by the magnetic field in the present model.   
In the weak coupling limit, typical quantities characterizing 
the phase diagram have the field-independent single energy scale, while  
its dimensionless coefficient 
varies with the magnetic field. 
The magnetic field-induced $d$FSD instability is 
a promising scenario for Sr$_{3}$Ru$_{2}$O$_{7}$ 
and we have discussed various future issues to establish such a scenario.

\begin{acknowledgments}
The author is grateful to A. A. Katanin for collaboration 
on a related work and fruitful discussions. 
He also thanks C. Honerkamp, G. Khaliullin, 
A. P. Mackenzie, W. Metzner, and R. S. Perry for helpful discussions, 
and R. Zeyher for critical reading of the manuscript. 

\end{acknowledgments}

\appendix
\section{Phase diagram for $\boldsymbol {\mu=-0.35}$ 
and $\boldsymbol {-0.34}$}
Figures~\ref{phase-m1.0}(a) and \ref{phase-m1.2} are two typical phase 
diagrams as a function of $h$ in the present model. 
These phase diagrams, however, are not connected smoothly 
by changing $\mu$. 
In fact, two other types of the phase diagram appear 
in a very limited $\mu$ region as shown in \fig{phase-other}. 
For $\mu=-0.35$ [\fig{phase-other}(a)], 
the dome-shaped transition line is realized on the 
relatively large $h$ side. 
In addition, the region surrounded by a first order transition line 
appear around $(h,T)=(0,0)$. 
At $T=0$, therefore, there is reentrant behavior as a function of $h$  
and the symmetric phase appears 
in the intermediate $h$ region ($0.12 \lesssim h \lesssim 0.14$). 
For $\mu=-0.34$ [\fig{phase-other}(b)], the overall shape of the 
phase diagram is the same as that shown in \fig{phase-m1.2}, but 
a first order line appears for high $T$ near $h=0$, 
accompanied by a tricritical point. 
As a result,  a first order transition happens as a function of $T$ 
in a sizable $h$ region. This is a very special 
case in our model since a first order transition as a function of $T$ 
is usually realized in a very limited $h$ region 
as seen in Figs.~\ref{phase-m1.0}(a) and \ref{phase-m1.2}. 

\begin{figure}
\centerline{\includegraphics[width=0.4\textwidth]{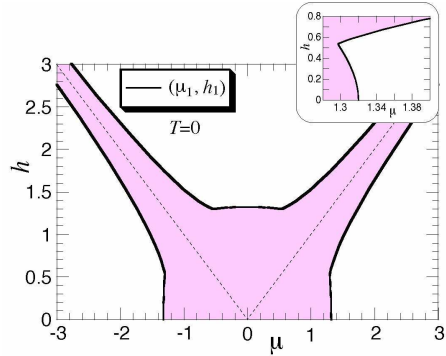}}
\caption{(Color online) The first order $d$FSD transition line 
in the weak coupling limit in the plane of 
$\mu$ and $h$ at $T=0$; $\mu$ and $h$ are scaled by 
the energy $\epsilon_{\Lambda} {\rm e}^{-1/(2\bar{g})}$; 
the symmetry-broken phase is stabilized in the colored (shaded) area; 
the dotted line represents the van Hove energy of 
the up-spin band ($\mu<0)$ and the down-spin band ($\mu>0$). 
The inset magnifies the region around $\mu\approx 1.3$.}
\label{phase-mh-weak}
\end{figure}

These peculiar types of the phase diagrams 
are understood from \fig{phase-mh-weak}(a) or similarly from 
\fig{phase-mh}(a). 
The first order transition line in \fig{phase-mh-weak}(a) is almost 
straight near 
$|\mu| \approx 1.3\epsilon_{\Lambda} {\rm e}^{-1/(2\bar{g})}$ 
[$\mu\approx -0.34$ in \fig{phase-mh}(b)]. 
Therefore we can have an extended $h$ region of the first order 
transition as seen in \fig{phase-other}(b). 
When we look at closely the region near 
$|\mu| \approx 1.3\epsilon_{\Lambda} {\rm e}^{-1/(2\bar{g})}$ 
and $h \lesssim 0.54\epsilon_{\Lambda} {\rm e}^{-1/(2\bar{g})}$ 
(inset of \fig{phase-mh-weak}), 
the first order transition line turns out to have 
a small inward curvature. 
This is why a symmetric phase is intervened between the two $d$FSD phases  
in \fig{phase-other}(a).

\section{Spin-dependent $\boldsymbol d$-wave Fermi surface deformation}
We have analyzed the $d$-wave Fermi surface symmetry breaking 
in the charge channel. From the point of view of Landau Fermi liquids, 
we can consider Fermi surface instability also in the spin channel. 
This possibility was pursued in several references\cite{varma06,quintanilla06,woelfle07,wu07} in the context of a general 
Landau Fermi liquid theory,\cite{quintanilla06,woelfle07,wu07} 
and a possible 
relation to a hidden order in URu$_{2}$Si$_{2}$.\cite{varma06} 
Here we clarify the relation between the charge-channel $d$FSD and 
the spin-channel $d$FSD, and discuss 
its relevance to Sr$_{3}$Ru$_{2}$O$_{7}$. 

The minimal model for the spin-dependent $d$FSD reads 
\be
 H = \sum_{\vk,\sigma} (\epsilon_{\vk}^{0}-\mu) \, n_{\vk \sigma} + 
 \frac{1}{2N} \sum_{\vk,\sigma,\vk',\sigma'} f_{\vk\vk'}^{a} \, 
\vS_{\vk}\cdot \vS_{\vk'}  
-h \sum_{\vk,\sigma}\sigma n_{\vk \sigma} \, ,
 \label{f+h-spin}
\ee
where $\vS_{\vk}=\frac{1}{2}\sum_{\alpha,\beta} 
c_{\vk \alpha}^{\dagger} {\boldsymbol \sigma}_{\alpha \beta} 
c_{\vk \beta}$, 
$f^{a}_{\vk \vk'}=-g^{a}d_{\vk}d_{\vk'}$, and the rest of notation 
is the same as the model (\ref{f+h-model}). 
Since the interaction has SU(2) symmetry and the magnetic field is 
assumed to be applied along the $z$ direction, we assume that the 
$S_{z}$ component can have a finite expectation value. 
Defining a mean field 
\be
\eta^{a}=-\frac{g^{a}}{N}\sum_{\vk}d_{\vk}\bra S_{\vk}^{z} \ket\,,
\ee
we obtain the mean-field Hamiltonian 
\be
H_{\rm MF} = \sum_{\vk,\sigma} \xi_{\vk \sigma}^{a} n_{\vk \sigma} 
+ \frac{N}{2 g^{a}}(\eta^{a})^{2}
\ee
where $\xi_{\vk\sigma}^{a}=\epsilon^{0}_{\vk} +\frac{1}{2}\sigma\eta^{a}d_{\vk}
-\mu_{\sigma}$. The grand canonical potential per lattice site thus reads 
\bea
\omega&=&-\frac{T}{N}\sum_{\vk,\sigma}\log 
(1+{\rm e}^{-\xi_{\vk \sigma}^{a}/T}) 
+\frac{(\eta^{a})^{2}}{2g^{a}}\\
&=&-\frac{T}{N}\sum_{\vk,\sigma}\log 
(1+{\rm e}^{-\xi_{\vk \sigma}^{a '}/T}) 
+\frac{(\eta^{a})^{2}}{2g^{a}} \,. 
\label{spin-freeenergy}
\eea
In the second line, we have introduced 
\be
\xi_{\vk\sigma}^{a '}=\epsilon^{0}_{\vk} +
\frac{1}{2}\eta^{a}d_{\vk}-\mu_{\sigma}\,,
\label{spin-dispersion}
\ee
noting that $d_{\vk}$ changes its sign 
with respect to $k_{x}\leftrightarrows k_{y}$ so that 
$\sigma d_{\vk}$ in the original $\xi_{\vk \sigma}^{a}$ can be 
written as $d_{\vk}$ in \eq{spin-dispersion}. 

\begin{figure}
\centerline{\includegraphics[width=0.9\textwidth]{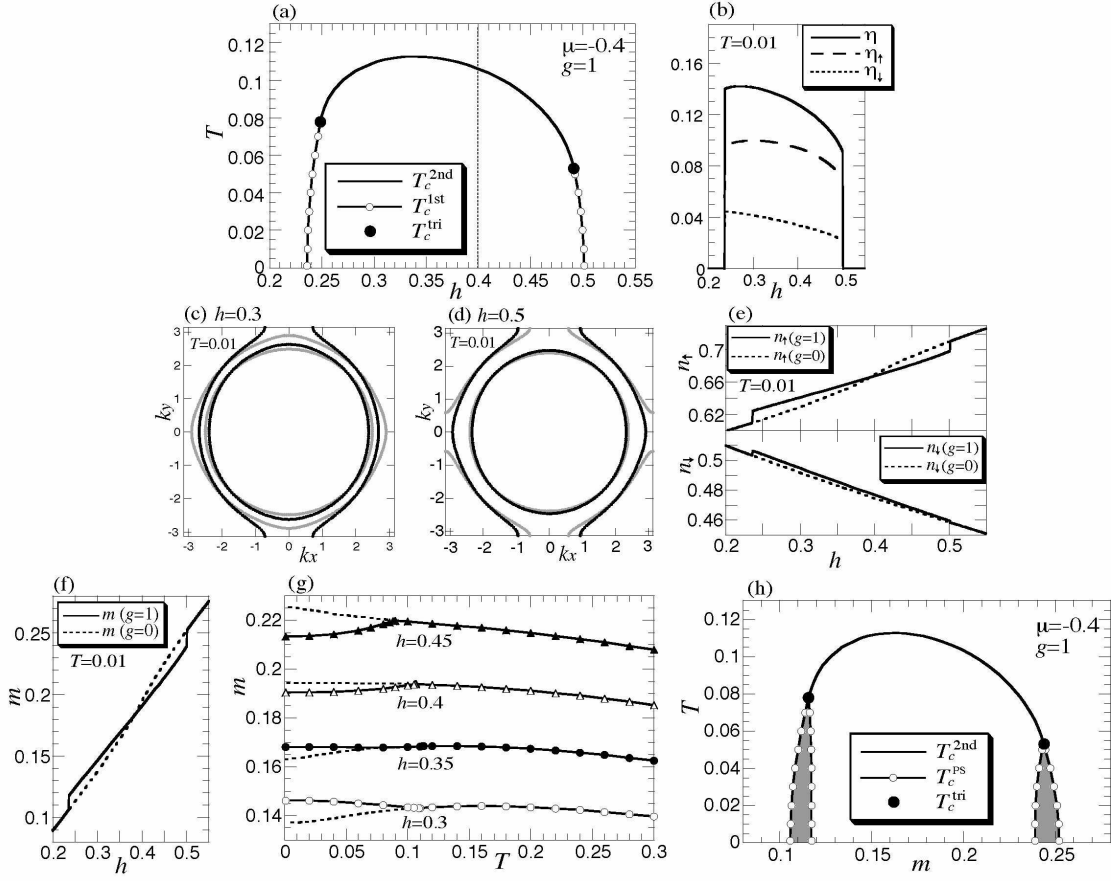}}
\caption{The mean-field solution for $t'/t=0.35$, $g=1$, and $\mu=-0.4$. 
(a) $h$-$T$ phase diagram; the transition line contains 
a second order line, $T_{c}^{\rm 2nd}$, for high $T$ and 
two first order line, $T_{c}^{\rm 1st}$, for low $T$; the solid circles 
are tricritical points; the dotted line denotes the van Hove energy 
of the up-spin band. 
(b) $h$ dependence of the order parameter at $T=0.01$; note that 
$\eta=\eta_{\uparrow}+\eta_{\downarrow}$. 
(c) and (d) FSs for $h=0.3$ and $0.5$ at $T=0.01$; the solid lines 
(gray lines) are  FSs for $g=1$ ($g=0$);  
the deformation of the inner FS in (d) is hardly  visible.  
(e) $h$ dependence of $n_{\sigma}$ at $T=0.01$ for $g=1$ 
(solid line) and $0$ (dotted  line). 
(f) $h$ dependence of the magnetization at $T=0.01$ for $g=1$ 
(solid line) and $0$ (dotted line). 
(g) $T$ dependence of the magnetization for several choices of $h$; 
the dotted line represents the result for $g=0$. 
(h) $m$-$T$ phase diagram; the system undergoes phase separation 
in the shaded regions surrounded by $T_{c}^{\rm PS}$; the other 
notation is the same as that in (a).}
\label{phase-m1.0}
\end{figure}

\begin{figure}
\centerline{\includegraphics[width=0.4\textwidth]{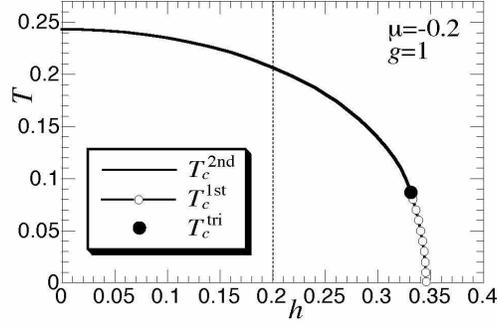}}
\caption{$h$-$T$ phase diagram for $t'/t=0.35$, $g=1$, and $\mu=-0.2$; 
the notation is the same as that in \fig{phase-m1.0}(a).}  
\label{phase-m1.2} 
\end{figure}

\begin{figure}
\centerline{\includegraphics[width=0.4\textwidth]{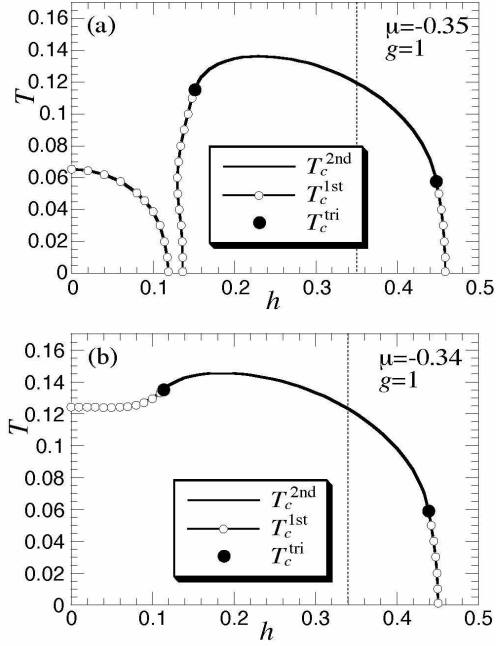}}
\caption{$h$-$T$ phase diagram for $\mu=-0.35$ (a) and $\mu=-0.34$ (b) 
for  $t'/t=0.35$ and $g=1$; 
the notation is the same as that in \fig{phase-m1.0}(a).}
\label{phase-other}
\end{figure}

\begin{figure}
\centerline{\includegraphics[width=0.4\textwidth]{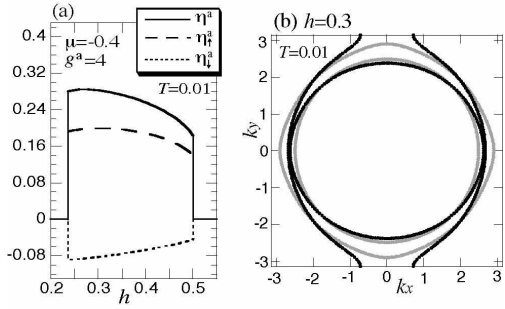}}
\caption{The mean-field solution in the spin-channel 
$d$FSD instability for  $t'/t=0.35$, $\mu=-0.4$, and $g^{a}=4$. 
(a) $h$ dependence of the order parameter at $T=0.01$. 
(b) FSs at $h=0.3$ (solid lines); the gray lines denote the FSs 
for $g^{a}=0$.}
\label{spin-dFSD}
\end{figure}

Comparing Eqs.~(\ref{spin-freeenergy})~and~(\ref{spin-dispersion}) with 
Eqs.~(\ref{freeenergy})~and~(\ref{charge-dispersion}), respectively, 
we see that 
the free energy becomes exactly the same under the following mapping, 
\be
\frac{1}{2}\eta^{a} \leftrightarrows \eta \,, \quad 
\frac{1}{4}g^{a} \leftrightarrows g \,.\label{map} 
\ee
Hence the thermodynamics in the spin channel of the $d$FSD is the 
same as that in the charge channel in the sense that 
we obtain the exactly the same results 
as Figs.~\ref{phase-m1.0}(a) and (e)-(h) under the mapping (\ref{map}).  
The difference appears in the \lq\lq internal'' 
structure of the order parameter and in a 
deformation of the FS. In the spin channel, we can write 
$\frac{1}{2}\eta^{a}=\frac{1}{2}\sum_{\sigma}\sigma\eta^{a}_{\sigma}$, 
where 
\bea
\frac{1}{2}\eta^{a}_{\sigma}&=&-\frac{g^{a}}{4N}\sum_{\vk}d_{\vk}
\bra n_{\vk \sigma} \ket \\
&=& \left\{ \begin{array}{l}
-\frac{g^{a}}{4N}\sum_{\vk}d_{\vk}f(\xi^{a '}_{\vk \uparrow}) \quad 
{\rm for\ } \sigma=\uparrow \\
+\frac{g^{a}}{4N}\sum_{\vk}d_{\vk}f(\xi^{a '}_{\vk \downarrow}) \quad 
{\rm for\ } \sigma=\downarrow \,.
\end{array}
\right.
\eea 
Comparing with \eq{charge-eta}, we see the relation under the mapping 
(\ref{map}) 
\be
\frac{1}{2}  \eta^{a}_{\uparrow}  \leftrightarrows \eta_{\uparrow} \,, 
\quad 
-\frac{1}{2} \eta^{a}_{\downarrow} \leftrightarrows \eta_{\downarrow}\,.
\ee
As we have actually seen in the numerical result in Sec.~III, 
$\eta_{\uparrow}$  in general has the same sign 
as $\eta_{\downarrow}$ in the charge channel. 
Therefore $\eta_{\uparrow}^{a}$ has an opposite sign of 
$\eta_{\downarrow}^{a}$ as seen in \fig{spin-dFSD}(a). 
While we have introduced $\xi_{\vk\sigma}^{a '}$ [\eq{spin-dispersion}], 
the Fermi surface itself is defined by $\xi_{\vk\sigma}^{a}=0$. 
Since $\xi_{\vk\sigma}^{a}$ contains a factor $\sigma\eta^{a}d_{\vk}$, 
a Fermi surface deformation in the spin channel 
occurs always in the opposite direction between the up-spin and 
the down-spin electron band as shown in \fig{spin-dFSD}(b); 
note that the deformation is determined by $\eta^{a}$, 
not by $\eta_{\sigma}^{a}$. 
As a result, the net deformation of the band is partially compensated.  
This is a crucial difference from 
the $d$FSD instability in the charge channel.

The recent experiment by Borzi {\it et al.}\cite{borzi07} 
showed a strong $xy$-anisotropy of the magnetoresistivity by applying 
an additional small magnetic field to the RuO$_{2}$ plane. 
This strong anisotropy may be discussed more naturally 
in terms of the $d$FSD 
instability in the charge channel rather than the spin channel. 
A conclusive discussion on which channel is more dominant would be 
to study microscopic deviation of the 
$d$FSD attractive interaction in both charge and spin channel 
in the context of Sr$_{3}$Ru$_{2}$O$_{7}$ 
and to compare the strength of the each channel.

\newpage
\bibliography{main.bib}

\begin{thebibliography}{48}
\bibitem[{mis()}]{misctJ}
\bibinfo{note}{The $d$FSD tendency in the $t$-$J$ model is observed in a rather
  wide doping region.\cite{yamase00,edegger06}}.
\end{thebibliography}

\newpage

\end{document}